\documentclass[11pt, a4paper]{article}
\pdfoutput=1
\usepackage{graphicx} % Required for inserting images
\usepackage{lineno}
\usepackage{amsmath}
\usepackage{multirow}
\usepackage{authblk}
\usepackage{setspace}

\usepackage{tabularx}
\usepackage{array}
\usepackage{authblk}
\usepackage{booktabs}
\usepackage{graphicx}
\usepackage{float}
\usepackage{subcaption}
\usepackage{amssymb}
\usepackage{lipsum} 
\usepackage{geometry}
\usepackage{hyperref}
\usepackage{color}
\title{ResoNet: Robust and Explainable ENSO Forecasts with Hybrid Convolution and Transformer Networks}

\author[1]{Pumeng Lyu}
\author[2]{Tao Tang}
\author[3]{Fenghua Ling}
\author[3]{Jing-Jia Luo}
\author[4]{Niklas Boers}
\author[1]{Wanli Ouyang}
\author[1,*]{Lei Bai}
\affil[1]{Shanghai Artificial Intelligence Laboratory, Shanghai, 200072, China}
\affil[2]{Zhejiang Meteorological Observatory, Hangzhou, 310002, China}
\affil[3]{Institute for Climate and Application Research (ICAR), Nanjing University of Information Science and Technology, Nanjing 210044, China}
\affil[4]{Technical University of Munich, Potsdam Institute for Climate Impact Research}
\date{\vspace{-5ex}} %% if you don't need date to appear
\affil[*]{Corresponding to \url{bailei@pjlab.org.cn}}

\begin{document}

\maketitle
\begin{abstract}
Recent studies have shown that deep learning (DL) models can skillfully predict the El Ni\~no-Southern Oscillation (ENSO) forecasts over 1.5 years ahead. However, concerns regarding the reliability of predictions made by DL methods persist, including potential overfitting issues and lack of interpretability. Here, we propose ResoNet, a DL model that combines convolutional neural network (CNN) and Transformer architectures. This hybrid architecture design enables our model to adequately capture local SSTA as well as long-range inter-basin interactions across oceans. We show that ResoNet can robustly predict ESNO at lead times between 19 and 26 months, thus outerforming existing approaches in terms of forecast horizon. According to an explainability method applied to ResoNet predictions of El Ni\~no and La Ni\~na events from 1- to 18-month lead, we find that it predicts the Ni\~no3.4 index based on multiple physically reasonable mechanisms, such as the Recharge Oscillator concept, Seasonal Footprint Mechanism, and Indian Ocean capacitor effect. Moreover, we demonstrate that for the first time the asymmetry between El Ni\~no and La Ni\~na development can be captured by ResoNet. Our results could help to alleviate skepticism about applying DL models for ENSO prediction and encourage more attempts to discover and predict climate phenomena using AI methods.
\end{abstract}

\doublespacing
\section*{Introduction}
The El Ni\~no-Southern Oscillation (ENSO), characterized by irregular oscillations between warm (El Ni\~no) and cold (La Ni\~na) phases, is one of the most pronounced inter-annual climate variability modes, exerting influence over global climate variations \cite{holton1989nino}. It has attracted great interest since the 1980s \cite{mccreary1981linear, philander1987simulation, neelin1992tropical}. Improvements in observing systems and ENSO prediction models help current statistical or dynamical models effectively predict El Ni\~no events with a notable lead time (i.e., 6 to 12 months) \cite{luo2008extended, barnston2012skill, ren2019seasonal, wang2020extended}. However, slow oscillating signals in ENSO, such as oceanic variations \cite{luo2008extended, gao2017roles}, equatorial winds \cite{gebbie2009predictability}, and sea surface temperature anomalies (SSTA) outside the equatorial Pacific \cite{park2018predicting}, suggest that there remains untapped potential to extend ENSO predictability to multiple years.

Deep learning methods have demonstrated remarkable advancements in various domains over the past decade \cite{lecun2015deep}. Inspired by the success of Convolutional Neural Networks (CNN) in computer vision \cite{krizhevsky2012imagenet, he2016deep}, Ham et al. (2019) used a three-layer CNN model and achieved effective ENSO forecasts 17 months ahead \cite{ham2019deep}. Since then, various kinds of deep learning models have been adopted in ENSO forecasts \cite{petersik2020probabilistic, cachay2020graph, yan2020temporal, gupta2020prediction, mu2022enso}. Currently, two major kinds of deep learning architectures are CNN and Transformers. CNN can efficiently handle datasets at different scales but has limitations in modeling long-range interactions \cite{dosovitskiy2020image}. Transformers, based on a self-attention-based architecture, can improve learning long-range interactions \cite{dosovitskiy2020image} and have already shown enhanced performances in weather forecasts \cite{bi2022pangu, chen2023fengwu}. Nevertheless, compared with CNNs, Transformers lack locality and translation equivariance \cite{dosovitskiy2020image}. Therefore, a larger number of training samples is required to effectively train a pure Transformer. However, climate data at a monthly scale is much smaller than weather data at an hourly scale. Even if we can use historical simulation data for training \cite{ham2019deep}, climate data is still insufficient compared with images in computer vision (usually over 100 million) \cite{dosovitskiy2020image}.Optmizing DL models with insufficient training samples is likely to cause overfitting issues, which hinders the application of Transformer-only models to capture climate dynamics like ENSO.

To address these challenges, we propose the \textbf{R}obust and \textbf{E}xplainable EN\textbf{SO} forecasting \textbf{Net}work (ResoNet). ResoNet is an integrated network that combines CNNs and Transformers. Combining CNNs and Transformers can improve model performances since this hybrid approach can adequately process local and global dynamics together \cite{dosovitskiy2020image, mehta2021mobilevit}. In addition, the bagging algorithm \cite{breiman1996bagging} is applied to train 20 models with different parameter initializations and training sets. With only three months' SSTA as model input, ResoNet on average can make effective predictions of Ni\~no3.4 index 18 months ahead in three test datasets we examined. We also emphasized the potential risks of overfitting issues. Under ideal settings, ResoNet can make effective forecasts up to 26 months in advance. However, this outstanding performance requires the over-ideal selection of model hyperparameters. 

Compared with the traditional dynamical or statistical methods, deep learning models for ENSO predictions are often doubted due to their lack of physical explanations. In this work, we employ the Integrated Gradient (IG) method \cite{sundararajan2017axiomatic} to explain the signals excavated by ResoNet to provide a physical intepretation. ResoNet can effectively capture the Recharge Oscillator paradigm \cite{jin1997equatorial, li1997phase}, the Seasonal Footprinting Mechanism \cite{vimont2003seasonal}, the inter-basin interactions among tropical oceans \cite{xie2009indian}, and the asymmetry of phase transitions between El Ni\~no and La Ni\~na events \cite{burgers1999normality, an2004nonlinearity}. These outcomes emphasize the model's capacity to capture multiple intricate dynamics within the climate system, providing valuable insights into the underlying mechanisms driving ENSO variation and predictability.

\section*{Results}
\subsection*{ResoNet Achieves Long-lead ENSO Predictions}
In this section, we present the quantitative prediction performance of ResoNet. Temporal Anomaly Correlation Coefficient (ACC) and Root Mean Square Error (RMSE) are used to validate the model performance. They are computed as a function of the forecast lead month t:

\begin{center}
\begin{equation}
    \begin{gathered}
        ACC(t) = \sum_{m=1}^{12} \frac{ \sum_{y=s}^{e}(O_{y,m}-\bar{O}_m)(F_{y,m,t}-\bar{F}_{m,t}) }{ \sqrt{ \sum_{y=s}^{e} (O_{y,m}-\bar{O}_m)^2 \sum_{y=s}^{e} (F_{y,m,t}-\bar{F}_{m,t})^2 } }
    \end{gathered}
\end{equation}

\begin{equation}
    \begin{gathered}
        RMSE(t) =  \frac{1}{12} \sum_{m=1}^{12} \sqrt{ \frac{1}{N} \sum_{y=1}^{N} (F_{y,m,t} - O_{y,m})^2 }
    \end{gathered}
\end{equation}
\end{center}

Here, O and F denote the observed and forecast values, respectively. $\Bar{O}_m$ and $\Bar{F}_{m,t}$ denote the temporal climatology concerning target season m (from 1 to 12) and the forecast lead months t (from 1 to 26). The label y denotes the forecast target year, respectively. 

\begin{figure}
\centering
    \includegraphics[width=\textwidth]{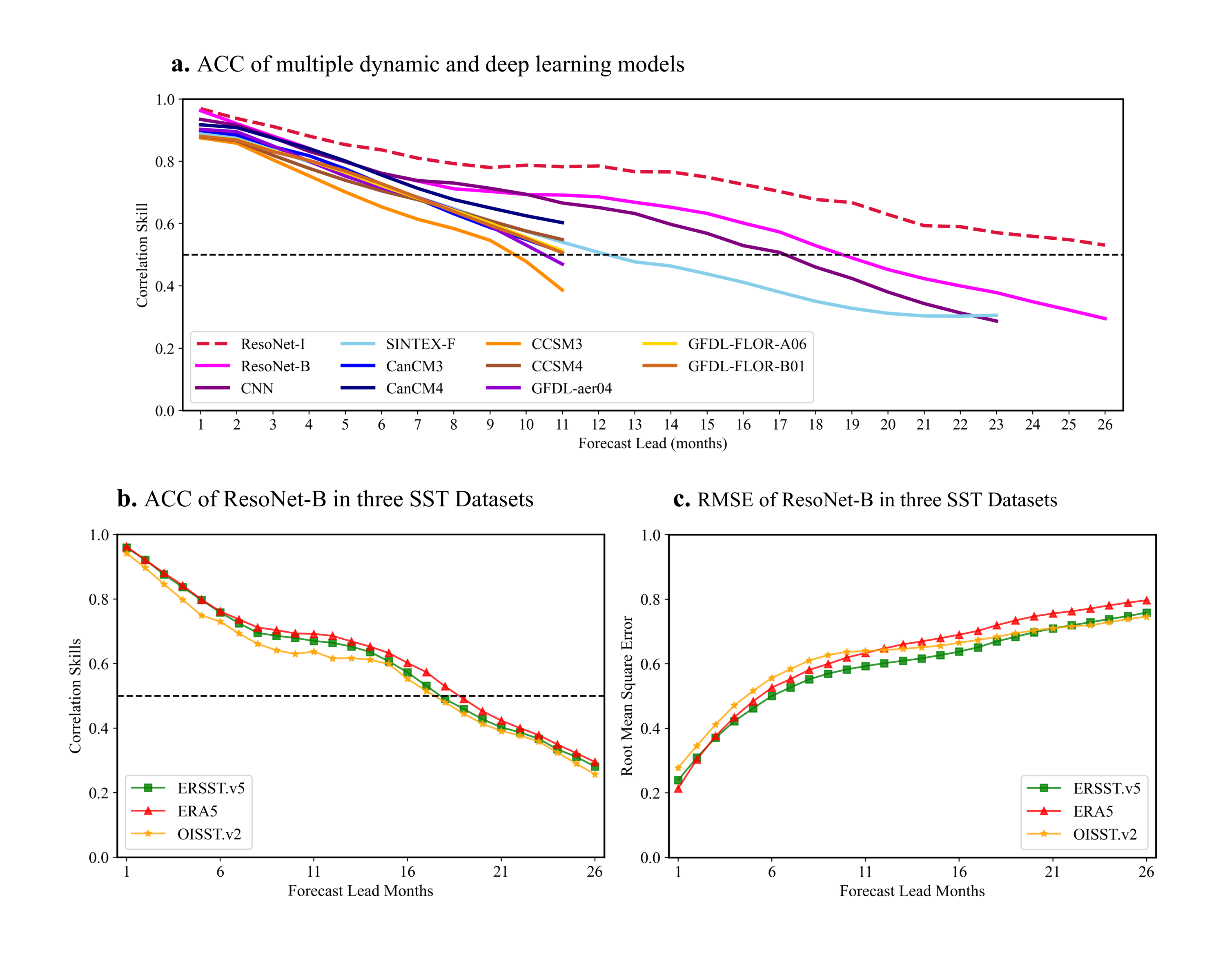}
\caption{Performances of ResoNet and other previous dynamical and deep learning models. (a) All-season ACC of ResoNet, CNN, SINTEX-F, and eight NMME models at different forecast lead months. (b) ACC of ResoNet using the bagging algorithm (ResoNet-B) at various lead times in three validation datasets: ERSST.v5, ERA5, OISST.v2. (c) Same as (b) but RMSE as performance metrics.}
\label{fig:1}
\end{figure}

To align with Ham et al.(2019) \cite{ham2019deep}, 32 years (1984-2017) SSTA from Extended Reconstructed Sea Surface Temperature, version 5 from the National Oceanic and Atmospheric Administration (ERSST.v5) were employed to examine and compare model performances. The all-season correlation skills for the three-month-running-averaged Ni\~no3.4 index from 1984 to 2017 of different models are shown in Figure. 1a. Here, ResoNet-B represents the ensemble mean predictions made from the 20 trained models using the bagging algorithm (see Methods). With the bagging algorithm, ResoNet models were trained 20 times separately using 20 different subsets of CMIP6 data, and transfer learning was then applied to 100 years of SSTA on ERSST.v5 (1871-1973). Compared with the CNN and dynamical models, ResoNet (ResoNet-B and ResoNet-I) shows substrantially improved prediction skill based on correlations, especially for long-term predictions with forecast lead beyond 11 months. The model exhibits all-season correlation skills above 0.5 for forecast lead up to 19 months. To examine whether ResoNet has overfitting problems, i.e., whether it is sensitive to different prediction test datasets or not, the correlation skills and root mean squared error based on two other datasets (see Methods) are also tested, i.e., ECMWF Reanalysis v5 (ERA5) \cite{hersbach2020era5} and NOAA Optimum Interpolation SST v2 (OISST.v2). ACC of ResoNet consistently exceeds 0.5 around 18 months in advance in all three datasets (see Fig. 1b), and the RMSE for the three different datasets (see Fig. 1c) does not vary greatly, suggesting that our model is reliable and robust in long-term ENSO forecasts. 

One advantage of using the bagging algorithm is that ensemble predictions do not rely on validation data. We would like to highlight that AI models might achieve exaggerated performance owing to their superior capability if trained with inappropriate settings, e.g., using the testing set in the hyperparameter selection stage. For example, if sorting 20 trained models and determining the ensemble of models from these 20 trained models based on the best performances on testing data, ResoNet could even achieve an effective forecast lead of 26 months ahead (see ResoNet-I in Fig. 1a and Supplementary Fig. 1). This results mean that our model can make reliable, effective 18 months forecast while could potentially make skillful predictions up to 26 months. However, testing data is always not available in real-time forecasts. Using testing data influences ensemble model decisions and leads to overly optimistic performance estimates since it can not generalize to new and unseen data. 

\subsection*{ResoNet Captures Explainable El Ni\~no Evolution Dynamics}
In addition to effective forecast skills, the physical mechanisms learned by ResoNet for El Ni\~no predictions can be unraveled. Correlations between input SSTA and the predicted Ni\~no3.4 index by ResoNet were computed at different forecast lead months (see Supplementary Fig. S3). Such linear regression method presents an overview of the linear dynamics ResoNet follows. To further explore regions that ResoNet identifies as responsible responsible for skillfully predicting El Ni\~no, an explainable deep learning method called Integrated Gradient (IG) was applied \cite{sundararajan2017axiomatic}. Attribution values were calculated by Equation \ref{eqn:ig}.

\begin{equation} \label{eqn:ig}
{IntegratedGrad}_i(x) = (x_i - {x_{i}}^{'}) \times \sum_{k=1}^{m} \frac{{\partial F(x^{'} + \frac{k}{m}(x - x^{'}))}}{{\partial x_i}} \times \frac{1}{m}
\end{equation}

Here, $m$ is the number of steps in the Riemann approximation of the integral. $i$ is the index of the input pixels to the model, which in this paper, i goes from 1 to n = 3 $\times$ 24 $\times$ 72 = 5,184. $x^{'}$ is the baseline we chose to compare outcomes. In this paper, the baseline is set as the zero-embedding vector with the same shape as the input to the model (3 $\times$ 24 $\times$ 72). For other details of IG, please refer to Sundararajan et al. (2017). Specifically, $IntegratedGrad_i(x)$ is the computed attribution value. The magnitude of attribution values can demonstrate which regions are sensitive and greatly affect the predictive skill of El Ni\~no index. In this paper, 800 attribution maps were computed for the 20 trained models for 40 years (1982-2021) of inputs in the DJF season. A sensitive region is defined where its attribution value during El Ni\~no years surpasses the 95\% confidence level, determined by comparing them to the averaged attribution values from 1982 to 2021. In this paper, ten El Ni\~no years (1983, 1987, 1988, 1992, 1995, 1998, 2003, 2007, 2010, 2016) between 1982 and 2021 were selected to analyze sensitive regions during El Ni\~no years.

\begin{figure}
\centering
    \includegraphics[width=\textwidth]{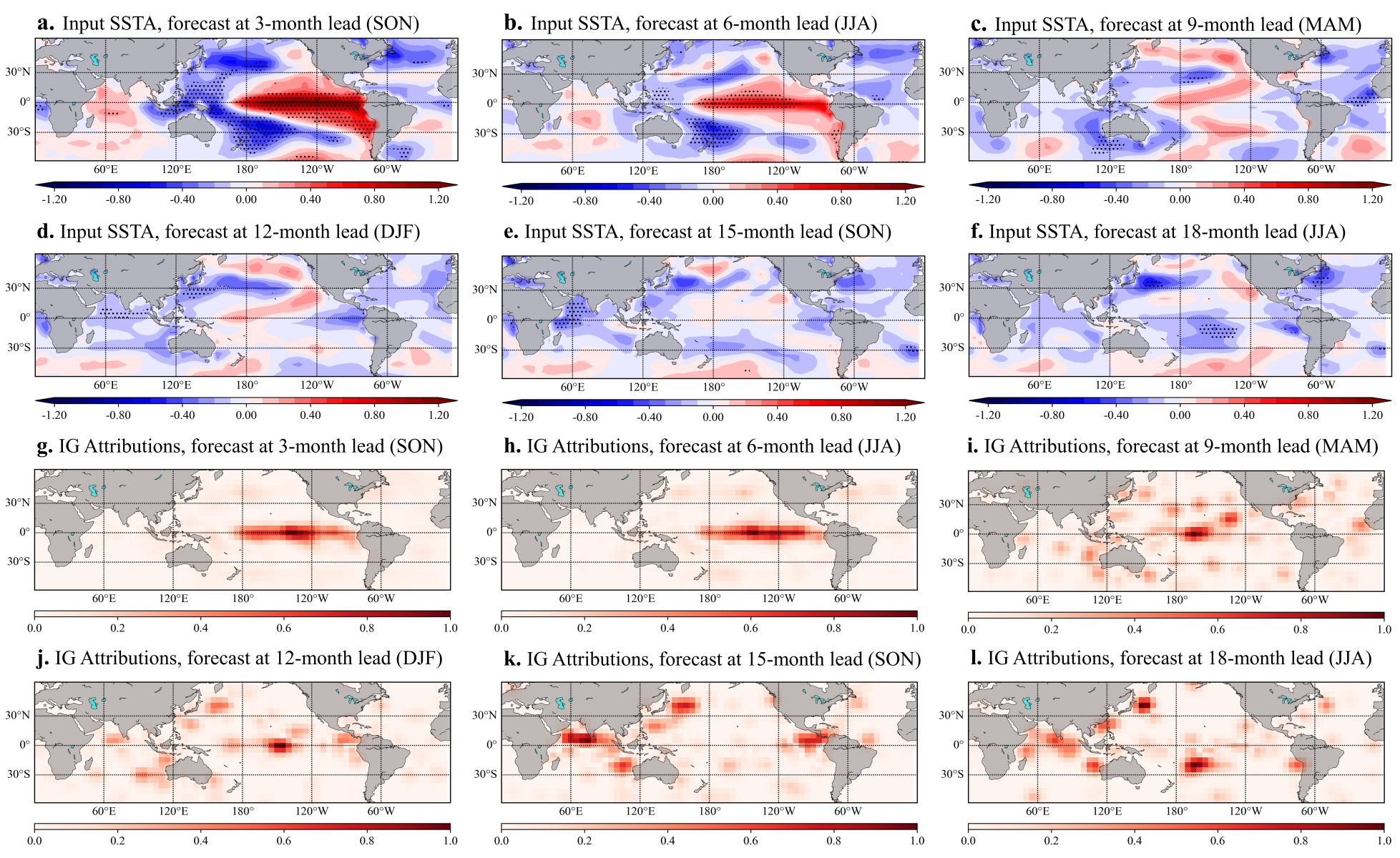}
\caption{Composite input SSTA and corresponding IG heatmaps in DJF season of 10 El Ni\~no events (1983, 1987, 1988, 1992, 1995, 1998, 2003, 2007, 2010, 2016), according to NOAA (\url{https://www.psl.noaa.gov/enso/}). (a)-(f) denotes the input SSTA with forecast lead at 3, 6, 9, 12, 15, and 18 months, respectively. Values over 95\% confidence level based on Student's \textit{t}-test are shaded. (g)-(l) denotes the attribution maps using the IG method with forecast lead at 3, 6, 9, 12, 15, and 18 months, respectively. All 20 trained models are considered for the attribution analysis. To avoid outliers that disrupt distributions of attribution values, attribution maps were first processed through a Gaussian filter with sigma 1.0. Only attribution values that surpass the 95\% confidence level are plotted.}
\label{fig:2}
\end{figure}

The advantage of using IG is that it can compare attribution values during El Ni\~no years and normal years. Composite input glkobal SSTA at different forecast leads to target DJF season were plotted (Fig. 2a-f), with significant regions over 95\% confidence level based on the Student \textit{t}-test shaded. Sensitive regions explored using IG on ResoNet align with significant regions in general, which demonstrates that ResoNet effectively captures sensitive regions that are known to play crucial roles in the formation of El Ni\~no events (see Fig. 2). When the forecast lead is within a year, attribution values are significantly located in the tropical Pacific Ocean (see Fig. 2g-j). However, as the forecast lead increases, ResoNet pays more attention to signals outside the equatorial Pacific (see Fig. 2k, l), such as signals in the tropical Indian Ocean, North Pacific Ocean, and South Pacific Ocean. This switching attention of local and global information is attributed to our model architecture, which combines CNN and Transformer.

Following the Recharge Oscillation mechanism \cite{jin1997equatorial}, an El Ni\~no event has been seeded according to the equatorward Sverdrup transport of subsurface warm water during phase transition from the cold to the warm phase \cite{zhang2019enso}. Cold SSTAs over the central-eastern equatorial Pacific keep the westerly wind anomaly, which further contributes to the poleward transport of surface cold water by Ekman transport. More specifically, at a lead of 18 months, negative correlations in the central and eastern tropical Pacific Ocean, along with positive correlations in the off-equatorial tropical and eastern Pacific Ocean (see Supplementary Fig. 3f), indicate the recharging process from the off-equatorial Pacific. This mechanism is well learned by ResoNet, as suggested by the significant attribution values over the off-equatorial Pacific. More evident signals appear over the southern Pacific than over the northern Pacific, which can be attributed to the attenuating effect of temperature anomalies in the northern Pacific \cite{schneider1999subduction}.

The equatorward transport of heat content leads to the deepening of the thermocline along the equator. Accordingly, the upwelling of the warm water over the central-eastern tropical Pacific gradually diminishes the cold SSTA there. The induced decrease of the trade wind over the tropical western Pacific would further result in the easterly expansion of the warm pool (see Fig. 2d, e). Correspondingly, the significant attribution values over the central Pacific imply the process above has been learned accurately by ResoNet (see Fig. 2j). When the forecast lead is within eight months, clear ENSO signals in the warm phase become evident (see Fig. 2a-c), the SSTA over the central-eastern tropical Pacific could grow up persistently according to the Bjerknes feedback \cite{bjerknes1969atmospheric}. Once again, the significant attribution values over the equator Pacific suggest that the key region of El Ni\~no development and the associated mechanism have been well captured by ResoNet (see Fig. 2g-i). 

Another key region captured by ResoNet is located in the northeastern Pacific with a one-year forecast lead (see Fig. 2j). The warm SSTA along the western coast of Northern America can propagate to the central equatorial Pacific and generate an El Ni\~no events around boreal spring according to the seasonal footprinting mechanism \cite{vimont2003seasonal}. Additionally, the tropical Indian Ocean is also considered by ResoNet as a key region with significant attribution values around 15 months ahead (see Fig. 2k). According to the Matsuno-Gill response \cite{matsuno1966quasi, gill1980some}, negative SSTA in the Indian Ocean induces divergence and westerly winds over the western tropical Pacific Ocean. Consequently, the induced eastward propagation of downwelling Kelvin waves potentially leads to the dissipation of La Ni\~na and initiation of El Ni\~no events, which is like the mechanism suggested by Xie et al. (2009)\cite{xie2009indian}. Results here demonstrate that ResoNet catches regional evolution dynamics in the tropical Pacific, but also long-range relationships outside the tropical Pacific, thanks to the hybrid model architecture combining CNNs and Transformers.

\subsection*{ResoNet Explores Asymmetric Behaviors between El Ni\~no and La Ni\~na}
One potential advantage of deep learning networks is their ability to build nonlinear and asymmetric relations. In addition to the analysis of the formation of El Ni\~no events, we also used the IG method to explore sensitive regions for La Ni\~na events at different forecast leads. Similar to the analysis of El Ni\~no events, seven La Ni\~na years (1989, 1999, 2000, 2008, 2011, 2012, 2021) were analyzed. Compared with El Ni\~no development, sensitive regions discovered during La Ni\~na development (see Fig. 3) are somewhat different from El Ni\~no development, especially when the forecast lead is 15-18 months, i.e., summer season a year before peak season. Results here indicate that apart from the linear mechanisms mentioned above, ResoNet could also catch the asymmetry between El Ni\~no and La Ni\~na development well.

\begin{figure}
\centering
    \includegraphics[width=\textwidth]{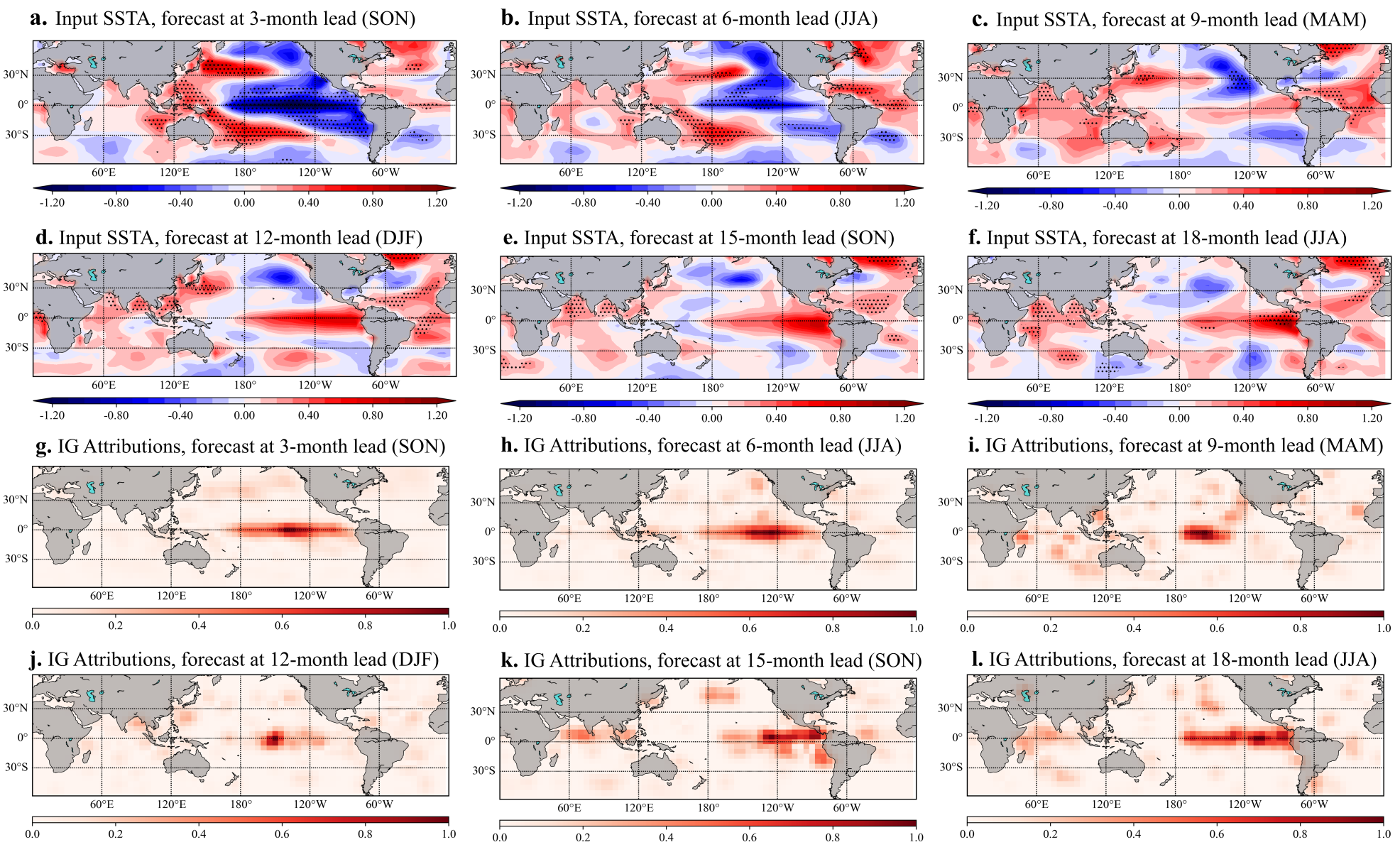}
\caption{Composite input SSTA and corresponding IG heatmaps in DJF season of 7 La Ni\~na events selected (1989, 1999, 2000, 2008, 2011, 2012, 2021), according to NOAA (\url{https://www.psl.noaa.gov/enso/}). (a)-(f) denotes the input SSTA with forecast lead at 3, 6, 9, 12, 15, and 18 months, respectively. Values over 95\% confidence level based on Student's \textit{t}-test are shaded. (g)-(l) denotes the attribution maps using the IG method with forecast lead at 3, 6, 9, 12, 15, and 18 months, respectively. All 20 trained models are considered for the attribution analysis. To avoid outliers that disrupt distributions of attribution values, attribution maps were first processed through a Gaussian filter with sigma 1.0. Only attribution values that surpass the 95\% confidence level are plotted.}
\label{fig:3}
\end{figure}

The physics behind the asymmetry between El Ni\~no and La Ni\~na development has been widely studied \cite{burgers1999normality, an2004nonlinearity}. However, this has never been demonstrated in existing works about applying AI methods in ENSO forecasts. More specifically, at the forecast lead of 18 months, while ResoNet suggests that the development of El Ni\~no is influenced by the southern Pacific (see Fig. 2l), it considers that La Ni\~na events are more sensitive to the eastern equatorial Pacific (see Fig. 3l). 

Based on the observational dataset (i.e., ERSST.v5), the asymmetry between the development of El Ni\~no and La Ni\~na can be demonstrated in more detail (see Fig. 4). Here, 40 years of Ni\~no 3.4 index in DJF season versus averaged SSTA in South Pacific and Ni\~no 3 region with forecast lead of 18 months are plotted. Linear regressions of 40 years of Ni\~no 3.4 index are shown in dashed green lines (see Fig. 3b and Fig. 3c). At forecast lead of 18 months, both SSTA in South Pacific and East tropical Pacific have a negative correlation to Ni\~no 3.4 index 18 months later in general. However, considering El Ni\~no and La Ni\~na years other than normal years, such correlations are somewhat asymmetric. 

With only 10 El Ni\~no and 7 La Ni\~na years between 1982 and 2021, we are mainly concerned with the location of quadrants for El Ni\~no or La Ni\~na years in scatter plots. Regarding the SSTA over the southern Pacific, almost all El Ni\~no events are located in the second quadrant, which implies the SSTAs over the southern Pacific in the previous summer have qualitatively consistent negative impacts on the El Ni\~no occur 18 months later. However, half of La Ni\~na events are evenly located in the third and fourth quadrants. Therefore, SSTAs over the southern Pacific have no qualitatively consistent impacts on the La Ni\~na events. The development of La Ni\~na is significantly correlated with the SSTAs over the Ni\~no 3 region 18 months ahead. In contrast, El Ni\~no events get no impacts from the SSTA averaged over Ni\~no3 region at 18 months ahead. This might be due to the slower demise of La Ni\~na than El Ni\~no as suggested by previous work \cite{chen2016relative}. El Ni\~no events have stronger amplitude but decay quickly, so SSTA at the tropical Pacific is more crucial for ESNO prediction during transitions from the warm phase to the cold phase in summer. Therefore, La Ni\~na events are predictable regarding signals at the tropical Pacific in the previous summer. However, stronger La Ni\~na events tend to last 2-3 years, so SSTA in tropical Pacific is not significant since it might complete transitions from cold to the warm phase or cold SSTA persist and another La Ni\~na event appears in the following year. The causes of this asymmetric evolution are due to the asymmetric SSTA pattern and nonlinear atmosphere responses. Therefore, different sensitive regions between El Ni\~no and La Ni\~na phase transitions suggest that ResoNet has correctly focused different regions according to different phase transitions, demonstrating its ability in nonlinear and asymmetric modeling.

\begin{figure}
\centering
\includegraphics[width=\textwidth]{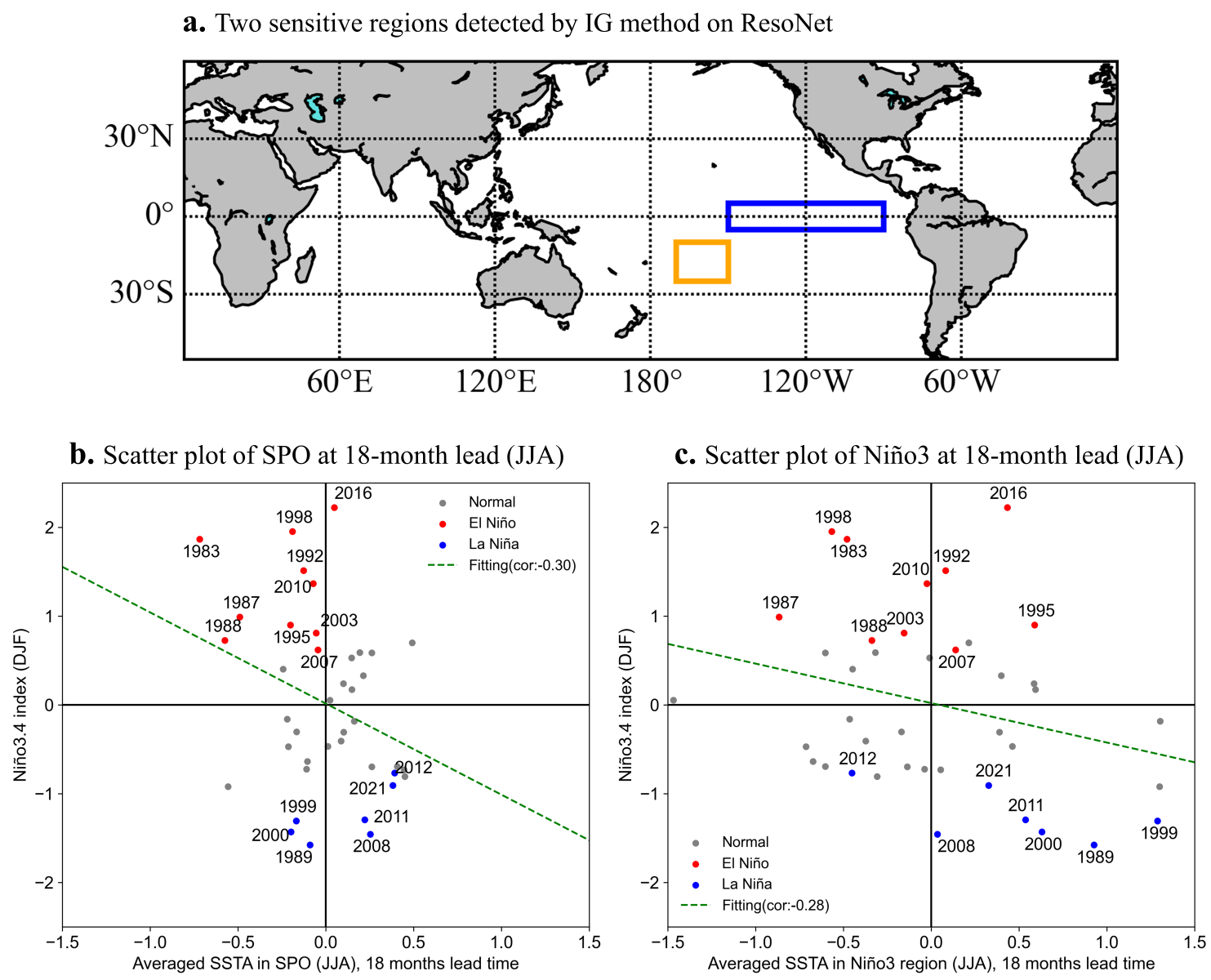}
\caption{Correlations between the SSTA over the southern Pacific, Ni\~no3 region at 18 months ahead and target Ni\~no3.4 index. (a) Locations of the southern Pacific (yellow, 25\textdegree - 10\textdegree S, 190\textdegree - 220 \textdegree E), and Ni\~no3 region (blue, 5\textdegree S - 5\textdegree N, 150\textdegree - 90 \textdegree W). (b) Scatter plot of averaged SSTA over the southern Pacific versus target Ni\~no3.4 index. (c) Same as (b) but for the SSTA over the tropical East Pacific Ocean (i.e., Ni\~no3 region).}
\label{fig:4}
\end{figure}

\section*{Discussion}
Recently, deep learning methods have been widely applied to climate forecasts to push the limits of prediction accuracy. Due to the lack of physical mechanisms, researchers might be concerned about whether such good performances with AI methods are reliable. By designing the novel CNN and Transformer hybrid model ResoNet and training with the bagging algorithm, our study here could avoid overfitting problems and enhance the performance, robustness, and interpretability of the ResoNet model. ResoNet could skillfully forecast ENSO 18 months ahead reasonably due to the excellent capturing of the recharge oscillator mechanism, the inter-basin interaction among tropical oceans, and the footprinting mechanism. In addition to improved prediction skills compared to existing approaches, there should be more analysis of the hidden dynamics AI models learn. For example, in this paper, our results demonstrate that ResoNet can forecast El Ni\~no and La Ni\~na events based on different key regions due to very good capturing of the asymmetry between the development of El Ni\~no and La Ni\~na events. This ability has never been achieved in previous AI models for ENSO forecasts. The development of explainable deep learning methods thus encourages discoveries of hidden physical mechanisms learned by deep learning models. While this study uses SSTA for model training, more complex implementations, including multiple variables, nonlinear interactions, and advanced explainable deep learning methods in climate forecasts, will be examined in future works. We hope our results can help alleviate the skepticism about applying Artificial Intelligence methods for ENSO prediction and support decision-making processes in various sectors that rely on accurate ENSO predictions. 

\section*{Methods}
\subsection*{Data}
Historical simulations produced by 20 models from Coupled Model Intercomparison Project phase 6 (CMIP6) were adopted to train ResoNet (see Supplementary Table S1). Only one member was selected for each CMIP6 model. Therefore, a total of 3,240 monthly samples from CMIP6 were used for training ResoNet, considering each target season and lead month. After training ResoNet with CMIP6 data, transfer learning was conducted with 103 years of reanalysis data (1871-1973) from the Extended Reconstructed Sea Surface Temperature, version 5 from the National Oceanic and Atmospheric Administration (ERSST.v5). Sea surface temperature data in 42 years (1980-2021) from NOAA Extended Reconstruction SSTs Version 5 (ERSST.v5), ECMWF Reanalysis v5 (ERA5), and NOAA Optimum Interpolation SST V2 (OISST.v2) were downloaded to validate the performance of ResoNet forecast. All data mentioned was interpolated into the regular grid (55\textdegree S - 60\textdegree N, 0\textdegree - 360 \textdegree E) with resolution 5\textdegree $\times$ 5\textdegree in both zonal and meridional direction. Predictions from 8 models of the North American Multi-model Ensemble (NMME) \cite{kirtman2014north} at 1-11 months lead and SINTEX-F \cite{luo2005reducing} at 1-23 months lead were collected to compare ResoNet model performance with previous dynamical models.

To process sea surface temperature data for model training and inference, original gridded data was first interpolated into the regular grid (55\textdegree S - 60\textdegree N, 0\textdegree - 360 \textdegree E) with resolution 5\textdegree $\times$ 5\textdegree in both zonal and meridional directions. Next, SSTA and the Ni\~no3.4 index were computed. Finally, SSTA was normalized by the spatially averaged standard deviation for easier model optimization. The number of forecast lead months is defined as the number of months between the latest input data and the middle month of the target season. The three-month-running mean of the Ni\~no3.4 index was the target output of ResoNet.

\subsection*{Architecture of ResoNet}
ResoNet (see Fig. 5) uses direct forecast strategy \cite{mu2022enso} for each target season and forecast lead. ResoNet consists of three parts: one embedding layer to process three-month SSTA inputs, three stage layers to extract local and global patterns, and one output layer to make predictions. Two main components in ResoNet are the Cross-scale Embedding Layer (CEL) and ENSO Mobile Block (EMB) in each stage layer. CELs in ResoNet are CNN-based layers that extract local features. It uses two different 2-dimensional convolutional kernels (2$\times$2 and 4$\times$4). Therefore, features at different scales and interactions between them can be extracted \cite{wang2018crossformer}. To reduce model inference cost and speed up feature extraction, the stride of each convolutional kernel in the CEL is set to 2$\times$2, resulting in a reduction of the spatial dimension by a factor of 4. 

ENSO Mobile Block (EMB) in ResoNet incorporates a self-attention-based Transformer between convolutional layers to explore global patterns \cite{mehta2021mobilevit}. Vision Transformer cuts images into patches and uses self-attention layers to process these patches as tokens. However, EMB uses Token Learner and Token Fuser to reduce model complexity and explore global relations effectively \cite{ryoo2021tokenlearner}. Token Learner processed 2-dimensional input features with shape $H \times W$ into S tokens (see Supplementary Fig. 1a). Then, Transformer is applied to extract interactions between these S tokens. Token Fuser thus projects these processed S tokens into shape $H \times W$ (see Supplementary Fig. 1b). A few convolutional layers are used to process local information and maintain data shape. By utilizing Token Learner and Token Fuser, EMB in ResoNet generates tokens for Transformer without the need to divide input data into non-overlapping patches. This patch-free design ensures that no information will lost when cutting non-overlapping patches. This hybrid and patch-free design of the ENSO Mobile Block improves the effectiveness of model training and enables the capturing of crucial global features for accurate ENSO forecasts. The detailed configuration of ResoNet is shown in Supplementary Table 2. 

\begin{figure}
\centering
\includegraphics[width=\textwidth]{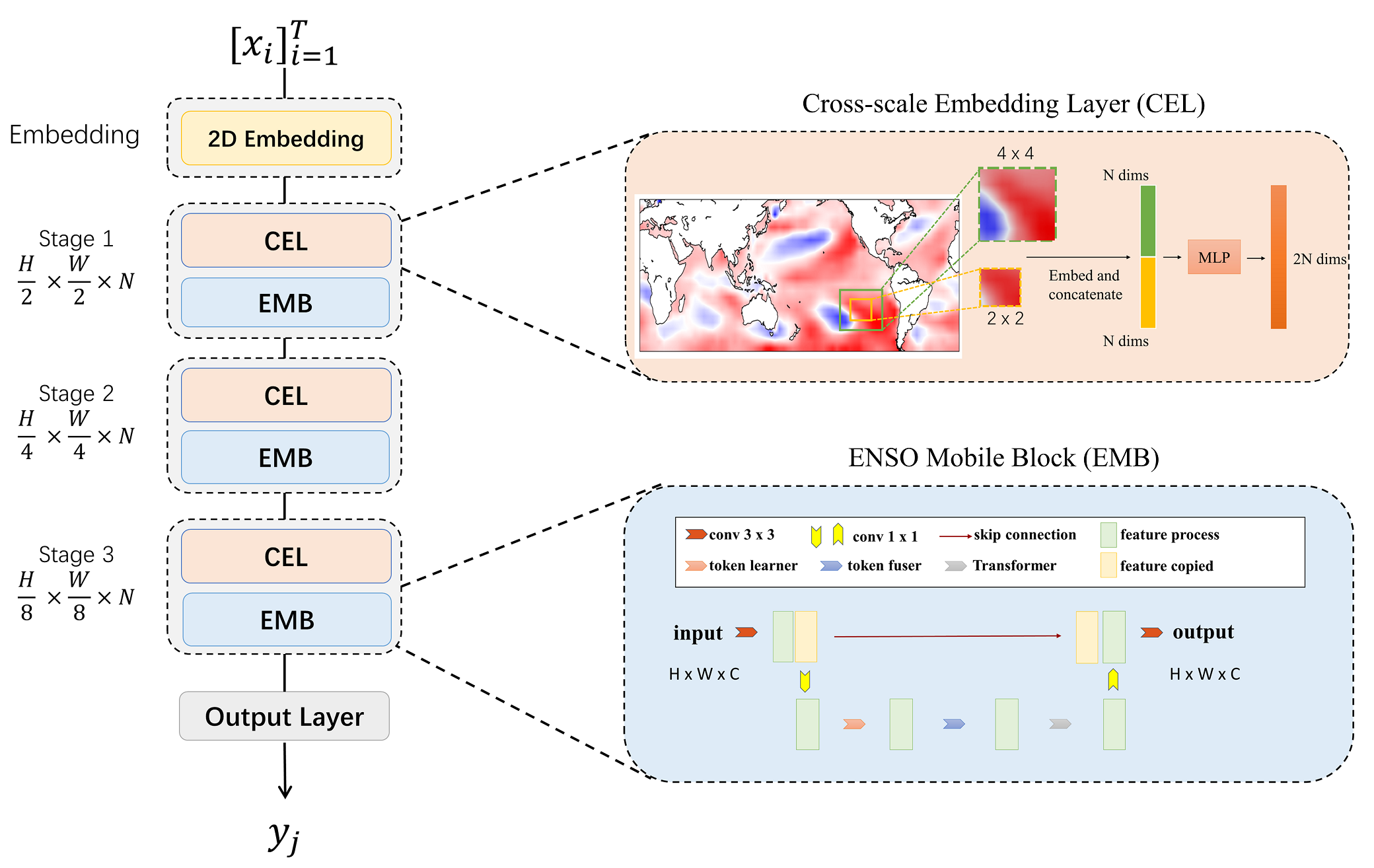}
\caption{Overall model architecture of ResoNet. ResoNet takes SSTA for three consecutive months as input. Global input SSTAs are embedded, concatenated, and mixed. Each stage layer consists of a Cross-Scale Embedding Layer (CEL), which is a CNN-based architecture that learns spatial local features and their interactions at different scales. Each CEL is followed by one ENSO mobile block (EMB), which is a Transformer-based architecture that captures global patterns that are crucial for ENSO forecast. After three stage layers, the output layer makes a three-month running mean of the Ni\~no3.4 index forecast. The right side of this figure shows detailed illustrations of CEL and EMB. Skip connections are used so processed information from both CNNs and Transformers can be kept for subsequent blocks.}
\label{fig:5}
\end{figure}

\subsection*{Bagging algorithm}
Because of a small number of training samples and variations of ENSO data, the bagging algorithm \cite{breiman1996bagging} was applied to reduce the instability of training and avoid overfitting. For each target season and lead month, 20 different training sets, each of size 3,240, were generated by sampling uniformly from 3,240 CMIP6 samples with replacement. Because every training set was sampled with replacement, they were independent of each other and each is expected to have the fraction (1 - 1/e)($\approx 63.2\%$) of 3,240 full CMIP6 samples ($\approx 2,050$ samples). The rest CMIP6 samples not selected were used as validation sets for early stopping during training. For each forecast lead month and target season, 20 different training sets from CMIP6 samples were generated to train 20 different ResoNet models with the same model structure but different model weights. Transfer learning was then applied to all 20 models with ERSST.v5 (1871-1973). Predictions were made by computing the ensemble mean of these 20 models.

\subsection*{Model training strategies}
ResoNet uses direct prediction strategy and predicts Ni\~no3. index. AdamW optimizer was used for training \cite{loshchilov2017decoupled}. To avoid gradient explosion and overfitting on training data, Smooth L1 Loss was used as the loss function for backward propagation \cite{girshick2015fast}. Equation \ref{eqn:smooth-l1-loss} gives the computation of Smooth L1 loss, with $x_n$ and $y_n$ denote the predictions and targets. Here, $\beta$ was set to be 0.5 for training ResoNet. Stochastic Gradient Descent with Restarts (SGDR) \cite{loshchilov2016sgdr} learning rate scheduler was used. The mini-batch size is 50 (20) for training on CMIP6 (ERSSTv5) samples. The learning rate is set to 5.0e-5.  Every training process stops when there is no improvement of Smooth L1 Loss on the validation set for 5 epochs. The maximum number of epochs for training with CMIP6 (ERSST.v5) samples is 100 (15). Detailed model training strategies are presented in Supplementary Table 3. 

\begin{equation} \label{eqn:smooth-l1-loss}
    l_n = \begin{cases}
    \frac{0.5(x_n - y_n)^2}{\beta}, & \text{if } |x_n - y_n| < \beta \\
    |x_n - y_n| - 0.5\cdot\beta, & \text{otherwise}
\end{cases}
\end{equation}

\section*{Data availability}
Data related to this paper can be downloaded from: \\ 
CMIP6 database, \url{https://esgf-node.llnl.gov/search/cmip6/}; \\
ERSST.v5 database, \url{https://www.ncei.noaa.gov/pub/data/cmb/ersst/v5/netcdf/}; \\
ERA5 database, \url{https://cds.climate.copernicus.eu/cdsapp#!/home}; \\
OISST.v2 database, \url{https://psl.noaa.gov/data/gridded/data.noaa.oisst.v2.html}, \\
NMME, \url{http://iridl.ldeo.columbia.edu/SOURCES/.Models/.NMME/}

\section*{Code availability}
The deep learning models were developed using standard libraries in open-source platforms including PyTorch (\url{https://pytorch.org/}). Codes used in this study are available from the corresponding author on request.

% \bibliographystyle{unsrt}
% \bibliography{refs}

\section*{Acknowledgements}
This work is supported by Shanghai Artificial Intelligence Laboratory and National Foundation of China (Grant 42030605).

\section*{Author contributions}
P.M.L. and T.T. are co-first authors. L.B. is the corresponding author who conceived the idea of this study. P.M.L and L.B. designed the AI models. P.M.L performed experiments. P.M.L. and T.T. performed the analysis and wrote the manuscript under the supervision of J.-J.L., N.B., L.B., and W.O.. All authors contributed to interpreting results, discussions of associated dynamics, and improvement of the presentation.

\section*{Competing interests}
All authors declare no competing interests.

\clearpage
\renewcommand{\figurename}{Supplementary Figure}
\renewcommand{\tablename}{Supplementary Table}

\subsection*{Supplementary Information}

\begin{table}[htbp]
  \centering
  \caption{CMIP6 models used for training}
  \begin{tabularx}{\textwidth}{>{\hsize=0.25\hsize\centering\arraybackslash}X>{\hsize=0.55\hsize\centering\arraybackslash}X>{\hsize=0.20\hsize\centering\arraybackslash}X}
    \toprule
    \textbf{Source ID} & \textbf{Modeling Groups} & \textbf{Integration Period} \\
    \midrule
    % Add your data rows here
    ACCESS-ESM-1-5 & Commonwealth Scientific and Industrial Research Organisation & \multirow{20}{*}{1850-2014} \\
    CanESM5 & Canadian Centre for Climate Modelling and Analysis & \\
    CESM2 & National Center for Atmospheric Research & \\
    CESM2-FV2 & National Center for Atmospheric Research & \\
    CESM2-WACCM & National Center for Atmospheric Research & \\
    CNRM6-CM6-1 & Centre National de Recherches Météorologiques & \\
    EC-Earth3-Veg & EC-Earth consortium & \\
    FGOALS-f3-L & Chinese Academy of Sciences & \\
    GISS-E2-1-G & NASA Goddard Institute for Space Studies & \\
    HadGEM3-GC31-LL & Met Office Hadley Centre & \\
    ICON-ESM-LR & Max Planck Institute for Meteorology & \\
    IPSL-CM6A-LR & Institute Pierre Simon Laplace & \\
    MCM-UA-1-0 & Department of Geosciences, University of Arizona & \\
    MIROC6 & Japan Agency for Marine-Earth Science and Technology & \\
    MPI-ESM-1-2-HAM & Max Planck Institute for Meteorology & \\
    MPI-ESM1-2-HR & Max Planck Institute for Meteorology & \\
    MPI-ESM1-2-LR & Max Planck Institute for Meteorology & \\
    MRI-ESM2-0 & Meteorological Research Institute & \\
    NESM3 & Nanjing University of Information Science and Technology & \\
    UKESM1-0-LL & National Institute of Meteorological Sciences/Korea Meteorological Administration & \\
    % Add more rows as needed
    \bottomrule
  \end{tabularx}
  \label{tab:cmip6-models}
\end{table}

\clearpage
\begin{table}[htbp]
  \centering
  \caption{Details of configurations of ResoNet.}
  \begin{tabularx}{\textwidth}{>{\hsize=0.25\hsize\centering\arraybackslash}X>{\hsize=0.35\hsize\centering\arraybackslash}X>{\hsize=0.20\hsize\centering\arraybackslash}X>{\hsize=0.20\hsize\centering\arraybackslash}X}
    \toprule
    \textbf{Module} & \textbf{Layer} & \textbf{Resolution} & \textbf{Channels}  \\
    \midrule
    \midrule
    Input & - & 24 $\times$ 72 & 1 \\
    \midrule
    \multirow{3}{*}{2d Embedding} & Conv3$\times$3 & 24 $\times$ 72 & 1 $\rightarrow$ 16\\
                                & GroupNorm & 24 $\times$ 72 & 16 \\
                                & LeakyReLU & 24 $\times$ 72 & 16 \\
    \midrule
    \multirow{1}{*}{1 \textsuperscript{st} CEL} & Conv2$\times$2, Conv4$\times$4 & 24 $\times$ 72 $\rightarrow$ 12 $\times$ 36 & 16 $\rightarrow$ 96 \\
    \midrule
    \multirow{7}{*}{1 \textsuperscript{st} EMB} & Conv3$\times$3, GroupNorm, SiLU & 12 $\times$ 36 & 96 \\
                                                                    & Conv1$\times$1, GroupNorm, SiLU & 12 $\times$ 36 & 96 \\
                                                                    & Token Learner & 12 $\times$ 36 $\rightarrow$ 2 & 3 $\times$ 96 \\
                                                                    & Transformer Layer $\times$3 & 2 & 3 $\times$ 96 \\
                                                                    & Token Fuser & 2 $\rightarrow$ 12 $\times$ 36 & 96 \\
                                                                    & Conv1$\times$1, GroupNorm, SiLU & 12 $\times$ 36 & 96 \\
                                                                    & Conv3$\times$3, GroupNorm, SiLU & 12 $\times$ 36 & 96 \\
    \midrule
    \multirow{1}{*}{2 \textsuperscript{nd} CEL} & Conv2$\times$2, Conv4$\times$4 & 12 $\times$ 36 $\rightarrow$ 6 $\times$ 18 & 96 \\
    \midrule
    \multirow{7}{*}{2 \textsuperscript{nd} EMB} & Conv3$\times$3, GroupNorm, SiLU & 6 $\times$ 18 & 96 \\
                                              & Conv1$\times$1, GroupNorm, SiLU & 6 $\times$ 18 & 96 \\
                                              & Token Learner & 6 $\times$ 18 $\rightarrow$ 4 & 3 $\times$ 96 \\
                                              & Transformer Layer $\times$3 & 4 & 3 $\times$ 96 \\
                                              & Token Fuser & 4 $\rightarrow$ 6 $\times$ 18 & 96 \\
                                              & Conv1$\times$1, GroupNorm, SiLU & 6 $\times$ 18 & 96 \\
                                              & Conv3$\times$3, GroupNorm, SiLU & 6 $\times$ 18 & 96 \\
    \midrule
    \multirow{1}{*}{3 \textsuperscript{rd} CEL} & Conv2$\times$2, Conv4$\times$4 & 6 $\times$ 18 $\rightarrow$ 3 $\times$ 9 & 96 \\
    \midrule
    \multirow{7}{*}{3 \textsuperscript{rd} EMB} & Conv3$\times$3, GroupNorm, SiLU & 3 $\times$ 9 & 96 \\
                                              & Conv1$\times$1, GroupNorm, SiLU & 3 $\times$ 9 & 96 \\
                                              & Token Learner & 3 $\times$ 9 $\rightarrow$ 4 & 3 $\times$ 96 \\
                                              & Transformer Layer $\times$3 & 4 & 3 $\times$ 96 \\
                                              & Token Fuser & 4 $\rightarrow$ 3 $\times$ 9 & 96 \\
                                              & Conv1$\times$1, GroupNorm, SiLU & 3 $\times$ 9 & 96 \\
                                              & Conv3$\times$3, GroupNorm, SiLU & 3 $\times$ 9 & 96 \\
    \midrule
    \multirow{4}{*}{Output Layer} & Global Average Pooling & 3 $\times$ 9 $\rightarrow$ 1 $\times$ 1 & 3 $\times$ 96 \\
                                  & Layer Normalization & 1 $\times$ 1 & 3 $\times$ 96 \\
                                  & Flatten & 1 $\times$ 1 & 3 $\times$ 96 $\rightarrow$ 288 \\
                                  & Fully Connected Layer & 1 $\times$ 1 & 288 $\rightarrow$ 1 \\
    \bottomrule
  \end{tabularx}
  \label{tab:resonet-config}
\end{table}

\clearpage
\begin{table}[htbp]
  \label{tab:S2}
  \centering
  \caption{ResoNet model training strategies.}
  \begin{tabular}{p{5cm} p{5cm} p{5cm}}
    \hline
    & \textbf{CMIP6 Training} & \textbf{Transfer Learning} \\
    \hline
    Batch Size & 50 & 20 \\
    Epochs & 100 & 15 \\
    Optimizer & AdamW & AdamW \\
    Initial LR & 5.0e-5 & 5.0e-5 \\
    Scheduler & SGDR & SGDR \\
    Weight Decay & 0.0001 & 0.0001 \\
    \hline
  \end{tabular}
\end{table}

\clearpage
\begin{figure}
    \centering
    \begin{align}
        \includegraphics[width=\textwidth]{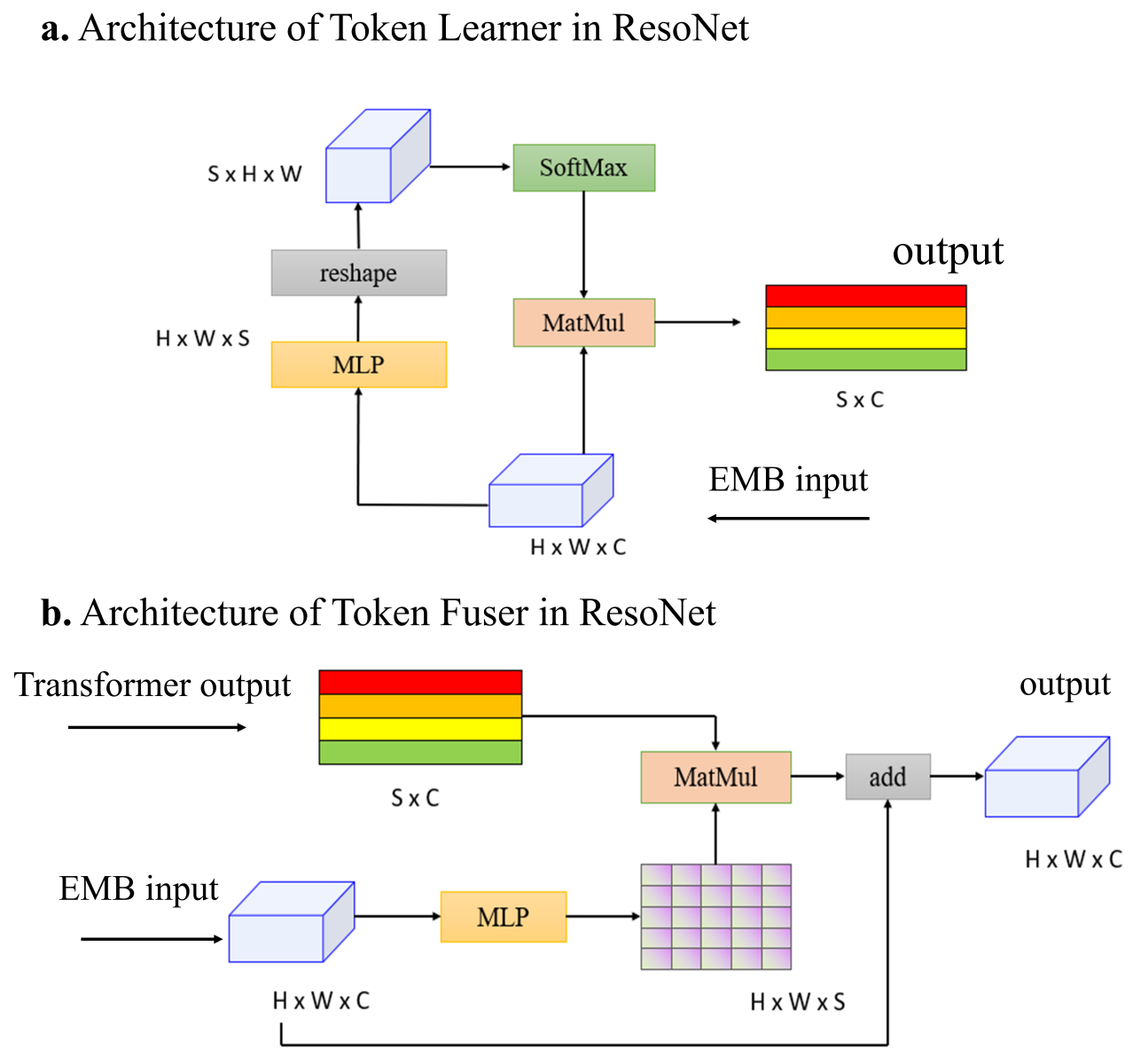}
    \end{align}
\caption{(a) Token Learner and (b) Token Fuser used in ENSO mobile block (EMB). With MLP layer and matrix multiplication, Token Learner transfers information from $H \times W$ grid points into S tokens. After Transformer layers, Token Fuser takes S tokens and input to Token Learner and projects feature shape back to $H \times W$ grid points. Here, C is the number of feature channels.}
\label{fig:s1}
\end{figure}

\clearpage
\begin{figure}[htbp]
\centering
\begin{align}
    \includegraphics[width=\textwidth]{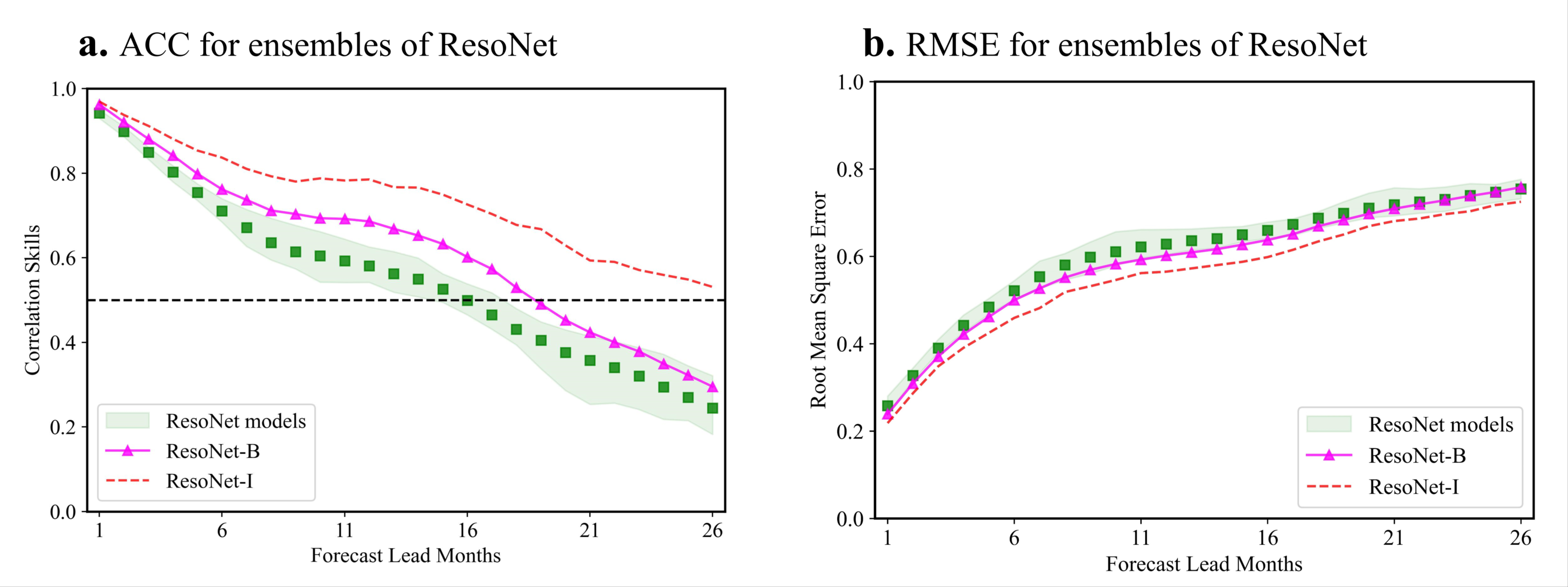}
\end{align}
\caption{Performances of ResoNet of 20 trained models and three ensemble modeling methods at forecast lead months from 1 to 26. (a) Correlations of all-season three-month averaged Ni\~no3.4 index at forecast lead from 1 to 26 months. Distributions of 20 single models are displayed as green-shaded regions. (b) Same as (a) but for root mean square error (RMSE).}
\label{fig:s2}
\end{figure}

\clearpage
\begin{figure}[htbp]
\begin{align}
    \includegraphics[width=\textwidth]{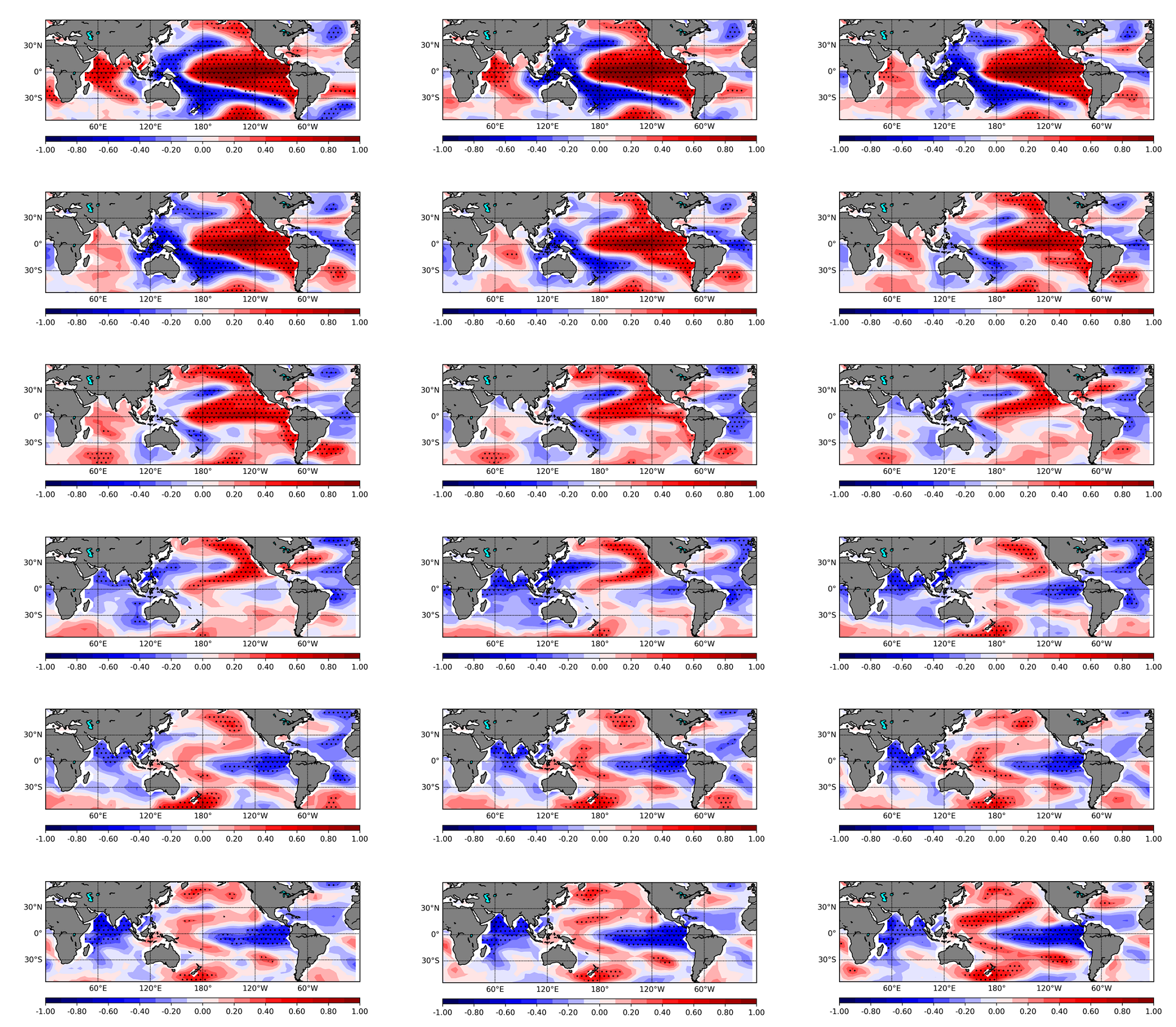}
\end{align}
\centering
\caption{Correlation between ResoNet's input and output at forecast lead from 1 to 18 months. Correlations are computed by linear regression of DJF season from 1982 to 2021. Significant regions (p $\leq$ 0.01) according to linear regression are shaded.}
\label{fig:s3}
\end{figure}

\clearpage
\begin{figure}[htbp]
\begin{align}
    \includegraphics[width=\textwidth]{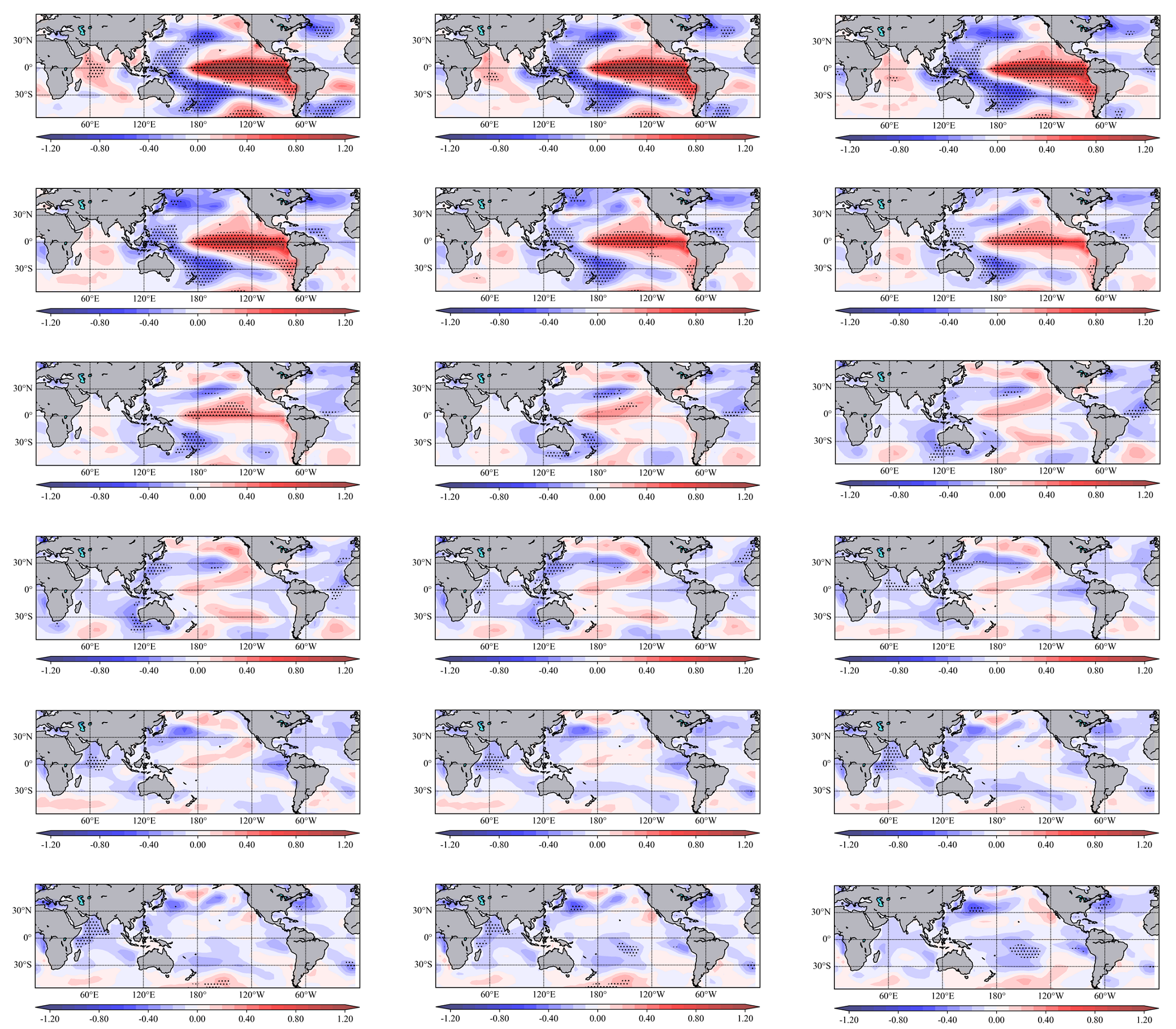}
\end{align}
\centering
\caption{Composite SSTA inputs of 10 El Ni\~no events at forecast lead from 1 to 18. Values over 95\% confidence level based on Student's \textit{t}-test are shaded.}
\label{fig:s4}
\end{figure}

\clearpage
\begin{figure}[htbp]
\begin{align}
    \includegraphics[width=\textwidth]{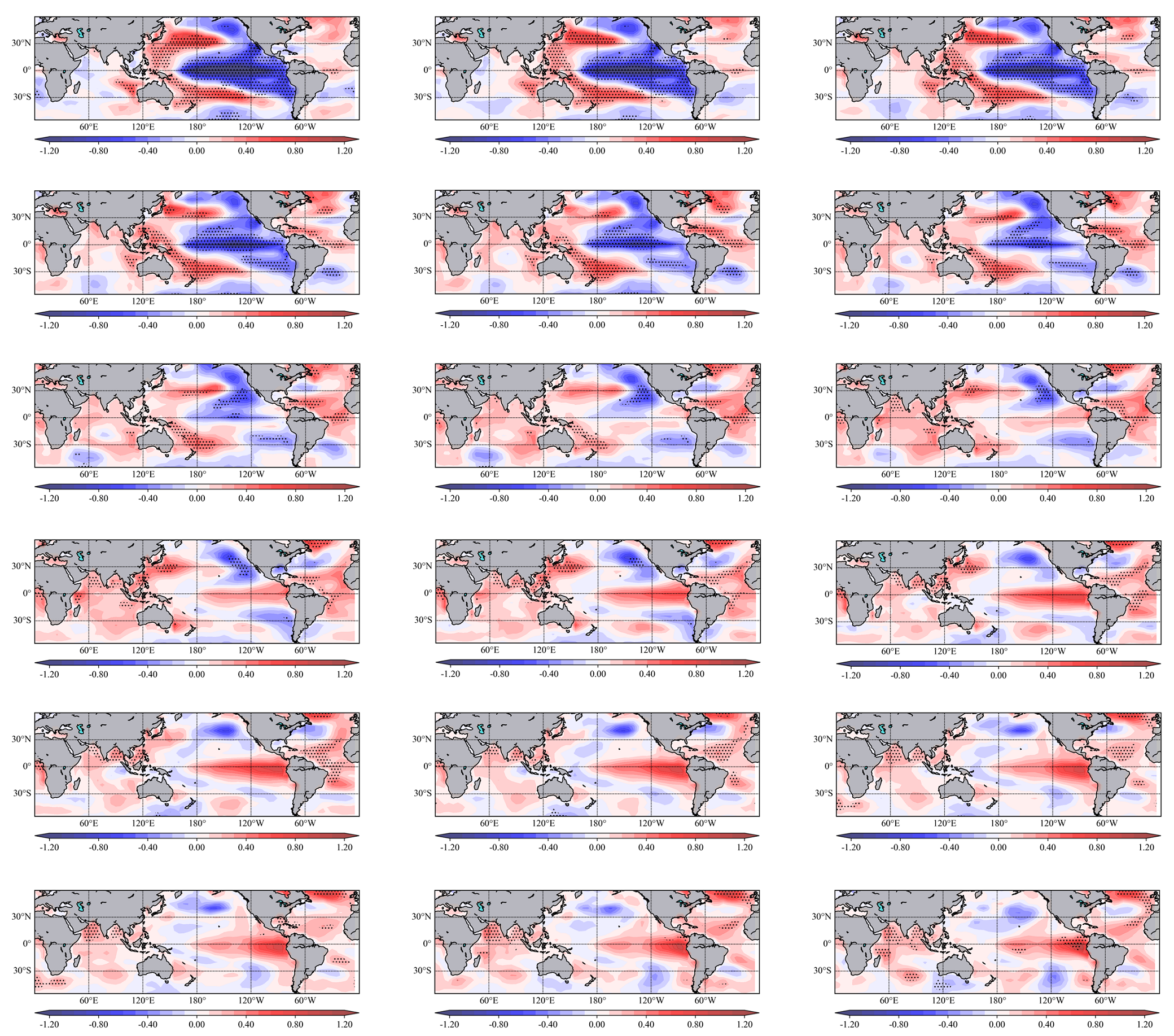}
\end{align}
\centering
\caption{Composite SSTA inputs of 7 La Ni\~na events at forecast lead from 1 to 18. Values over 95\% confidence level based on Student's \textit{t}-test are shaded.}
\label{fig:s5}
\end{figure}

\clearpage
\begin{figure}[htbp]
\begin{align}
    \includegraphics[width=\textwidth]{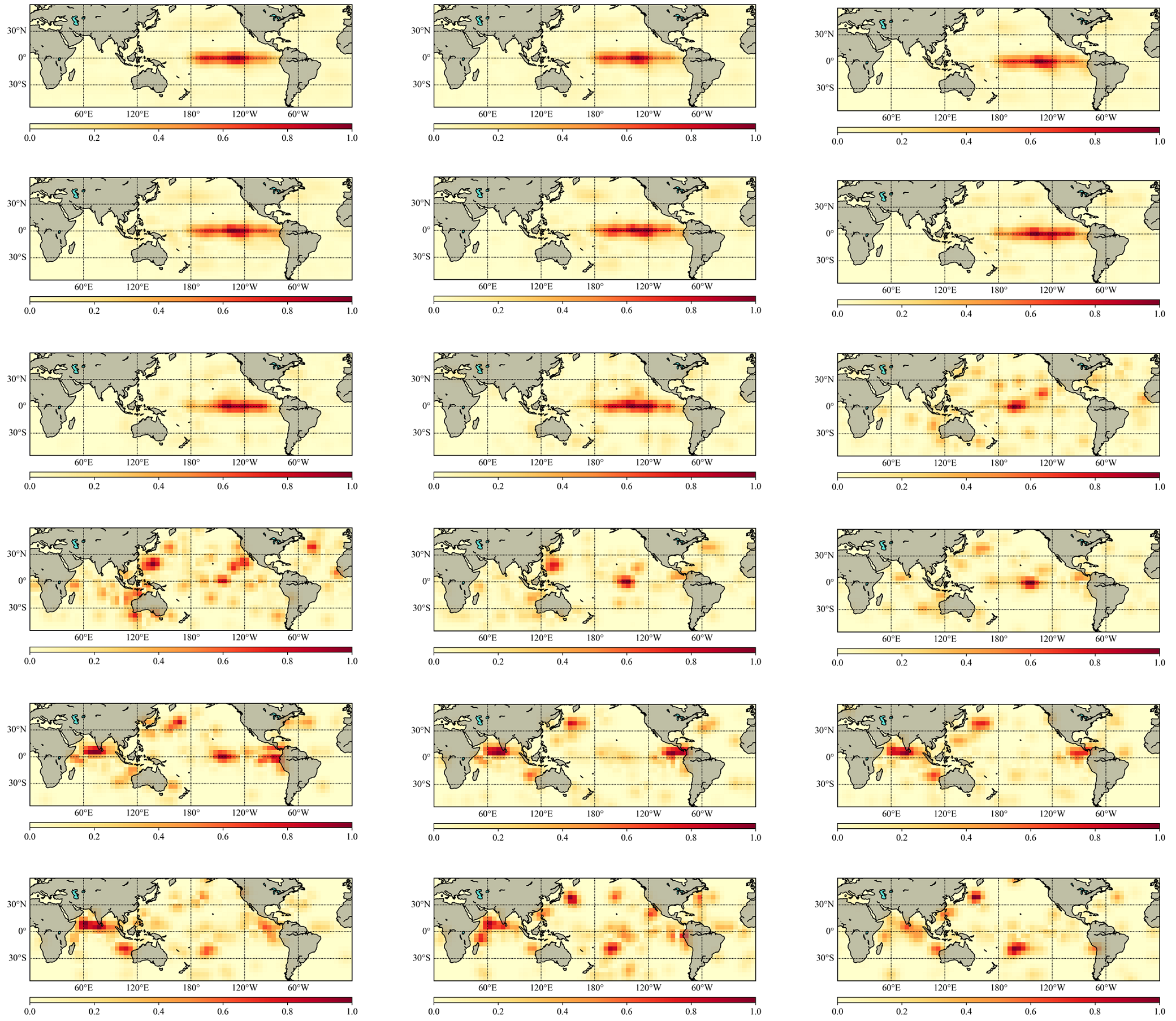}
\end{align}
\caption{Attribution maps of 10 El Ni\~no events in DJF Season at forecast lead months from 1 to 18. All 20 trained models are considered for the attribution analysis. Only attribution values that surpass the 95\% confidence level, determined by comparing them to the averaged attribution values from 1982 to 2021, are plotted. }
\label{fig:s6}
\end{figure}

\clearpage
\begin{figure}[htbp]
\begin{align}
    \includegraphics[width=\textwidth]{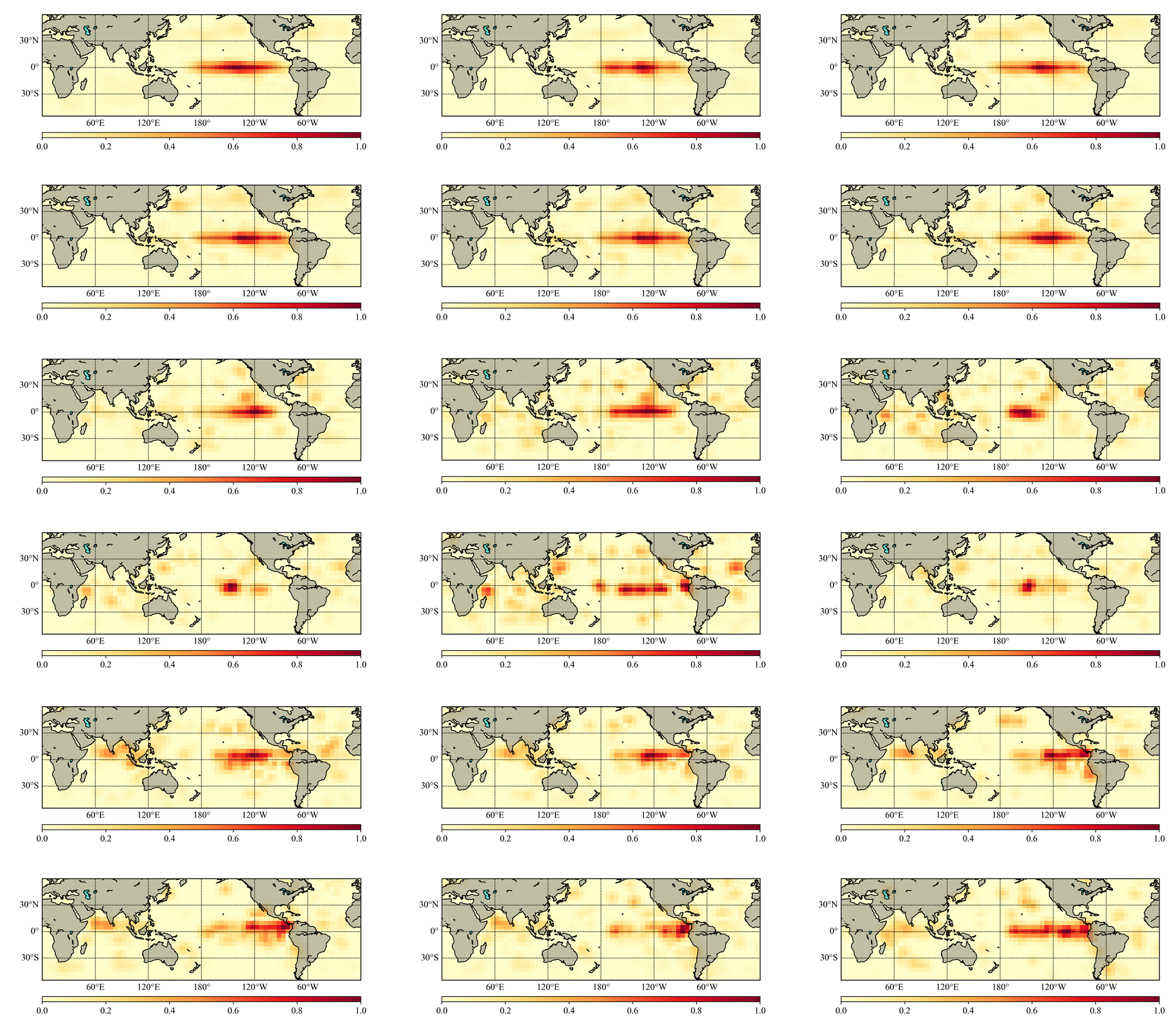}
\end{align}
\centering
\caption{Attribution maps of 7 La Ni\~na events in DJF Season at forecast lead months from 1 to 18. All 20 trained models are considered for the attribution analysis. Only attribution values that surpass the 95\% confidence level, determined by comparing them to the averaged attribution values from 1982 to 2021, are plotted. }
\label{fig:s7}
\end{figure}

\end{document}